\documentclass[%
 reprint,
superscriptaddress,
 amsmath,amssymb,
 aps,
pra,
longbibliography,
]{revtex4-2}

\usepackage{graphicx}
\usepackage{dcolumn}
\usepackage{bm}

\usepackage{blindtext}
\usepackage{siunitx}
\DeclareSIUnit\gauss{G}
\newcommand{\LEmuSR}{LE-$\mathrm{\mu}$SR}

\DeclareSIUnit\eVperc{\eV\per\clight}
\DeclareSIUnit\clight{\text{\ensuremath{c}}}

\usepackage{lineno}

\begin{document}

\preprint{APS/123-QED}

\title{Monitoring the tagging efficiency of the Low-Energy Muon beamline through background analysis: Insights into the long-term performance of ultrathin carbon foils}
\author{Gianluca Janka}
\email{gianluca.janka@psi.ch}
\affiliation{PSI Center for Neutron and Muon Sciences CNM, 5232 Villigen PSI, Switzerland}

\author{Herman Ishchenko}
\affiliation{PSI Center for Neutron and Muon Sciences CNM, 5232 Villigen PSI, Switzerland}
\affiliation{Department of Physics, University of Basel, Klingelbergstrasse 82, CH-4056 Basel, Switzerland}

\author{Zaher Salman}
\affiliation{PSI Center for Neutron and Muon Sciences CNM, 5232 Villigen PSI, Switzerland}

\author{Andreas Suter}
\affiliation{PSI Center for Neutron and Muon Sciences CNM, 5232 Villigen PSI, Switzerland}

\author{Thomas Prokscha}
\email{thomas.prokscha@psi.ch}
\affiliation{PSI Center for Neutron and Muon Sciences CNM, 5232 Villigen PSI, Switzerland}
\date{\today}

\begin{abstract}
The efficient tagging of individual muons, along with the stability of the involved ultrathin carbon foil, is critical for ensuring fast, reliable and reproducible low-energy muon spin rotation (\LEmuSR) measurements. At the Paul Scherrer Institute’s Low-Energy Muon (LEM) beamline, we developed a method to monitor the tagging efficiency of the beamline using routinely collected muon decay histograms. This method leverages background comparison before and after the arrival time of the muons at the sample to extract the tagging efficiency, eliminating the need for additional detectors or measurements. By analyzing data collected between 2018 and 2024, we confirm the method's reliability and validate its results using independent reference measurements. Furthermore, we establish a correlation between the tagging efficiency and foil thickness by investigating the impact of contamination and outgassing on the carbon foil, as well as the restoration effects of laser cleaning. The findings underscore the importance of monitoring the carbon foil's condition to maintain consistent beamline performance and reproducibility. This method offers a practical approach for detector efficiency monitoring, applicable to other beamlines employing similar setups.

\end{abstract}
\maketitle


\section{\label{sec:Intro}Introduction}

\begin{figure*}[t!]
    \centering
	  	\centering
	    \includegraphics[width=0.95\textwidth, trim={0 0 0 0},clip]{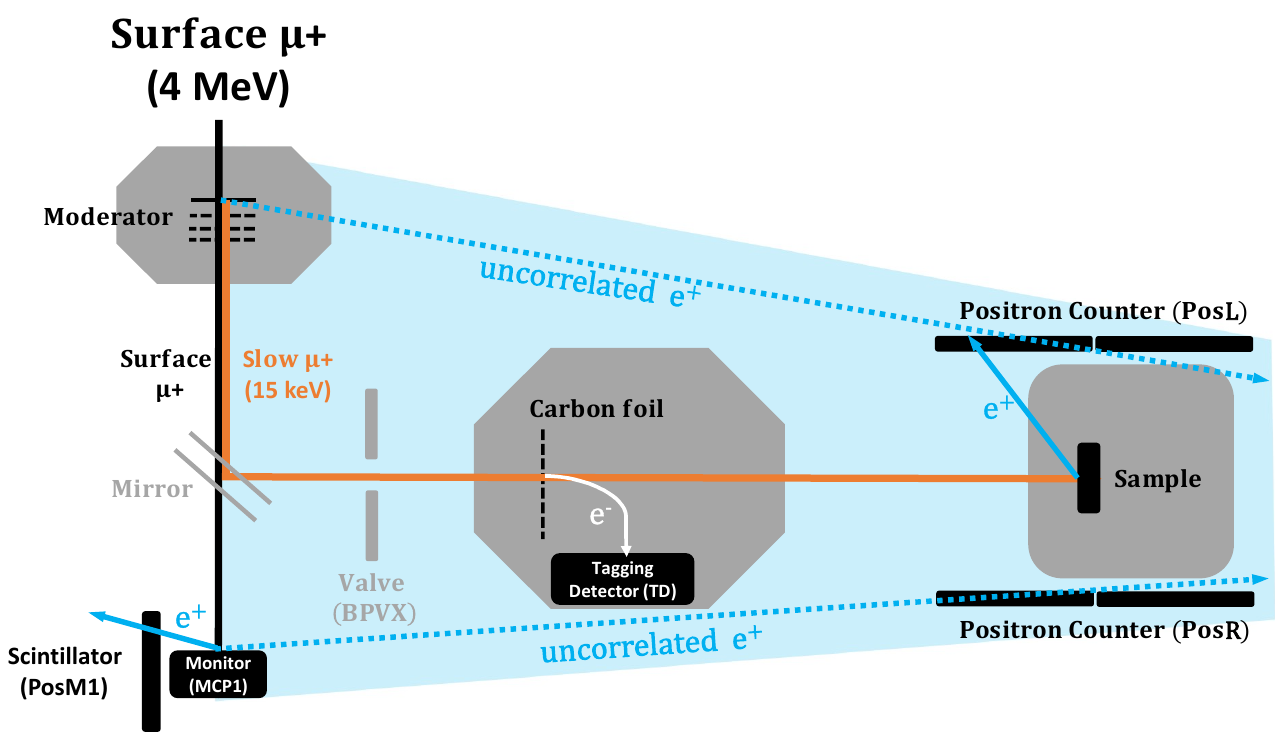}
  
	  \caption[]{\label{fig:lem} Simplified schematic showing the top view of the LEM beamline. The uncorrelated e$^+$ background is depicted as a light-blue shaded region, with common pathways represented by dashed lines. The trajectory of the low-energy muons is shown as an orange line, ending on the sample. The path of fast surface muons is illustrated with a black line, stopping on the beam monitor MCP1.}
\end{figure*}

The Low-Energy Muon (LEM) beamline (see Fig.~\ref{fig:lem}) receives a continuous beam of over \SI[per-mode=symbol]{4e8}{\per\second} spin-polarized, positively charged surface muons ($\mu^+$) with a momentum of \SI[per-mode=symbol]{28}{\mega\eVperc} from the $\mu$E4 beamline at the Paul Scherrer Institute (PSI) \cite{2008_Prokscha}.
These muons are focused onto a moderation target consisting of a \SI{125}{\micro\meter} thick silver foil, cooled to approximately \SI{10}{\kelvin}, and coated with a thin (several hundred \si{\nano\meter}) layer of a solid rare gas, such as argon or neon \cite{2001_Prokscha}. 
Approximately half of the surface muons stop and decay in the silver foil. A small fraction (\numrange{5e-5}{e-4} \cite{2004_Morenzoni}) of $\mu^+$ particles are moderated in the solid rare gas layers to \si{\electronvolt} energies, preserving their polarization \cite{1994_Morenzoni}, and re-accelerated through a set of grids into a monoenergetic low-energy $\mu^+$ beam of up to \SI{20}{\kilo\electronvolt}. 
To separate the low-energy $\mu^+$ beam from the remaining surface muons, an electrostatic mirror tuned to the energy of the low-energy $\mu^+$ deflects them in the horizontal plane by \SI{90}{\degree}, while the surface muons remain unaffected. These surface muons continue their straight path, stopping and decaying in a beam monitor (MCP1) at the end of the beamline, where a set of scintillators (PosM1) is positioned as an additional monitor for the decay positrons.

The continuous nature of the low-energy beam means that there is no inherent starting time for muons passing through the beamline. 
To address this, a tagging setup consisting of an ultrathin (sub \SI{10}{\nano\meter}) carbon foil and a microchannel plate detector (trigger detector, TD) was implemented \cite{1998_Hofer, 2015_Khaw}. 
When a $\mu^+$ passes through the carbon foil, it interacts with the material, loses \SIrange{5}{10}{\percent} of its energy, and releases secondary electrons \cite{2024_Janka}. The yield of secondary electrons strongly depends on the surface cleanliness of the carbon foils \cite{2016_Allegrini,2003_Allegrini, 1996_Rothard}. These secondary electrons are guided onto the TD and generate the trigger signal for the next measurement; the $\mu^+$ has been tagged (in experimental jargon).

The low-energy muons are eventually implanted into the sample. The most common measurements performed at the LEM beamline are either temperature or implantation depth scans conducted in a weak transverse magnetic field of \SIrange{50}{100}{\gauss}. The magnetic field is oriented parallel to the incoming beam, but perpendicular to the muon spin. The muon spin starts precessing around the magnetic field orientation. The $\mu^+$ is unstable and decays with a lifetime of \SI{2.2}{\micro\second} into a positron and two neutrinos. Due to parity violation in the muon decay, the positron is preferentially emitted in the direction of the muon spin. Detecting the emission direction of the positron provides information on the muon's spin polarization as a function of time and, therefore, on the local magnetic and electronic properties at the muon's stopping site (see Refs.~\cite{2022_Blundell, 2024_Amato} for more information about the muon spin rotation ($\mu$SR) technique). To detect the positrons from the muon decay, the sample is surrounded by positron counters at the top, bottom, left, and right (PosT, PosB, PosL, PosR) with respect to the incoming muon beam. A typical decay histogram recorded with PosL is shown in Fig.~\ref{fig:decay_histo}.

For the low-energy $\mu$SR (\LEmuSR) measurements, accurate knowledge of the implantation energy and beam spot size is crucial to ensure reproducible results. Additionally, since the beam spot size is directly influenced by multiple scattering in the carbon foil, the foil should be as thin as possible to minimize energy loss and beam straggling. All these parameters depend on a stable carbon foil \cite{2024_Janka}, meaning that the foil must maintain a consistent thickness over long periods. 
Over time, we observe that the carbon foil thickness increases due to residual gas accumulation or outgassing from samples. 
In addition, prolonged usage of a carbon foil can lead to the formation of holes, reducing the effective coverage and consequently lowering the tagging efficiency.
Thus, a method is required to monitor the condition of the carbon foil and enable timely intervention in case of significant changes.

In this work, we present such a method based on comparing the background levels of \LEmuSR~decay histograms before and after the time a muon reaches the sample, to estimate the tagging efficiency of the TD, which is closely related to the thickness of the carbon foil. As such, the tagging efficiency not only influences counting statistics but also serves as a sensitive proxy for both the condition of the carbon foil and the overall beam quality.
The required data for this method are routinely collected at the LEM beamline, enabling the extraction of the tagging efficiency as a byproduct, without the need for additional detectors or measurements.
We compare the results of this method with those obtained using an additional detector replacing the sample. We show tagging efficiency trends for the years 2018 to 2024 and demonstrate the efficacy of cleaning the carbon foil by laser illumination.


\section{\label{sec:TagEff}Extraction of tagging efficiency}

\begin{figure}[t!]
	  \centering
	\includegraphics[width=0.99\columnwidth, trim={0 0 0 0},clip]{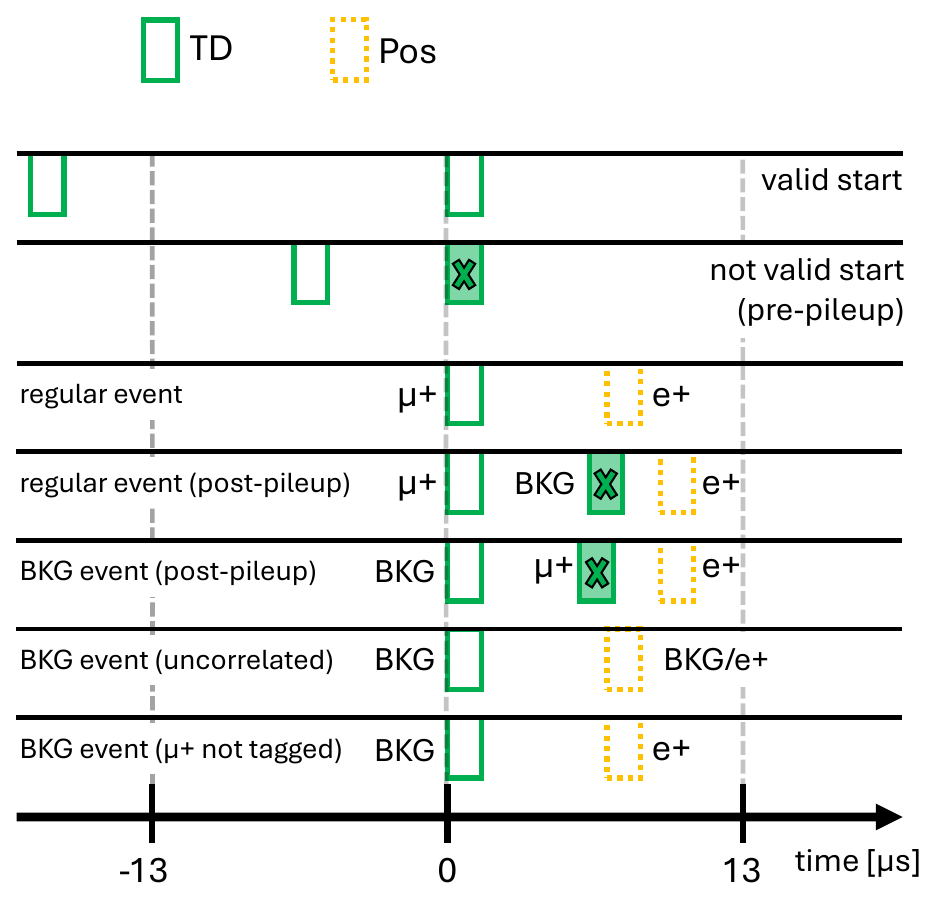} 
	  \caption[]{\label{fig:td_events}Examples of possible event sequences recorded in LE-$\mu$SR measurements. Hits on the TD are marked as solid green boxes, while hits on the positron counters are shown as dashed orange boxes. The type of particle responsible for each signal is indicated next to the corresponding box.}
\end{figure}

The method is based on counting the decay positrons ($e^+$) with the positron counters surrounding the sample region, in coincidence with a start signal generated by the TD. 
As in typical $\mu$SR measurements, a valid start by an incoming muon is defined by a TD event that has not been preceded by another TD event for at least the duration of the data gate (\SI{13}{\micro\second} in our case), see Fig.~\ref{fig:td_events}. If a TD event occurs within the previous \SI{13}{\micro\second} (pre-pileup), this start event is not valid and is discarded by the trigger logic.

In \LEmuSR~measurements, events with additional hits on the TD after the initial start time (post-pileup) are still recorded. This results in different background levels before and after the muon reaches the sample at the time $t =t_0$. 
Consequently, for background correction of the $\mu$SR histograms, the background cannot be determined from the $t<t_0$ region, as is typically done in $\mu$SR, but must instead be included in the fit for $t>t_0$.

Comparing the background levels before and after $t_0$ allows us to extract the tagging efficiency $\epsilon_\mathrm{TD}$ of the TD, as described in Sec.~\ref{sec:bkg_levels}. Figure~\ref{fig:td_events} shows examples of pulse sequences of events that can occur in a \LEmuSR~measurement. 
In addition to the histograms containing the post-pileup events, we also provide the standard $\mu$SR histograms in the \LEmuSR~data files, 
in which post-pileup events are excluded (post-pileup corrected histograms, or ppc histograms). In these histograms, the background level at $t < t_0$ is equal to the background level at $t > t_0$.

\subsection{Background levels comparison}\label{sec:bkg_levels}
The positron background of a raw \LEmuSR~ histogram before $t_0$ ($N_\mathrm{BKG}^{(t<t_0)}$) can be described as: 
\begin{align}
    N_\mathrm{BKG}^{(t<t_0)} \propto \underbrace{r_{e^+_\mathrm{uncorr}}}_\mathrm{uncorrelated} + \underbrace{(1-\epsilon_\mathrm{TD})\cdot r_{e^+_\mu}}_{\mu^+ \mathrm{not~tagged}}.
\end{align}
The first term, labeled $r_{e^+_\mathrm{uncorr}}$, represents the contribution from the rate of uncorrelated $e^+$ originating from scattered beam positrons, for example coming from the moderator target or the beam monitor MCP1. Muons decaying in-flight present a negligible contribution to the background. 
The second term accounts for correlated background events, where the $\mu^+$ was not detected by the TD, but a different background event triggered the TD and provided the trigger signal. The $\mu^+$ nonetheless reached the sample and decayed, and therefore its decay positron generated a signal in a positron counter. $r_{e^+_\mu}$ is the detected rate of $e^+$ originating from $\mu^+$ decaying in the sample region.

The positron background after the $t_0$, $N_\mathrm{BKG}^{(t>t_0)}$, is similar; it includes the uncorrelated rate $r_{e^+_\mathrm{uncorr}}$ and the rate of untagged $\mu^+$, as both rates are independent of the exact timing of the tagging detector relative to positron detection. 
However, an additional term must be included to account for post-pileup events. This term describes cases where a $\mu^+$ is detected by the TD, but the initial trigger signal was generated by a different background event such as an ion or beam positron. The full background $N_\mathrm{BKG}^{(t>t_0)}$ can be written as:
\begin{align}
    N_\mathrm{BKG}^{(t>t_0)} & \propto \underbrace{r_{e^+_\mathrm{uncorr}}}_\mathrm{uncorrelated} + \underbrace{(1-\epsilon_\mathrm{TD})\cdot r_{e^+_\mu}}_{\mu^+ \mathrm{not~tagged}} + \underbrace{\epsilon_\mathrm{TD} \cdot r_{e^+_\mu}}_\mathrm{post-pileup} \\
    & \propto r_\mathrm{LEM},
\end{align}
which is also proportional to the total detected positron rate $r_\mathrm{LEM}$:
\begin{align}
\label{eq:lem}
   r_\mathrm{LEM} & = r_{e^+_\mathrm{uncorr}} + r_{e^+_\mu}
\end{align}
The ratio $\alpha$ between the two background levels is given by:
\begin{equation}
    \alpha = \frac{N_\mathrm{BKG}^{(t>t_0)}}{N_\mathrm{BKG}^{(t<t_0)}} = \frac{r_\mathrm{LEM}}{r_{e^+_\mathrm{uncorr}} + (1-\epsilon_\mathrm{TD})\cdot r_{e^+_\mu}},
\end{equation}
which can be solved for the tagging efficiency $\epsilon_\mathrm{TD}$:

\begin{align}
 \label{eq:6}\epsilon_\mathrm{TD} &= \frac{\alpha(r_{e^+_\mathrm{uncorr}}+  r_{e^+_\mu}) - r_\mathrm{LEM}}{\alpha\cdot r_{e^+_\mu}} \\
\label{eq:TDEff}&= \frac{(\alpha - 1)\cdot r_\mathrm{LEM}}{\alpha(r_\mathrm{LEM}-r_{e^+_\mathrm{uncorr}})},
\end{align}
where we used the relation from Eq.~\ref{eq:lem} in the step from Eq.~\ref{eq:6} to Eq.~\ref{eq:TDEff}. 

Eq.~\ref{eq:TDEff} demonstrates that by determining the background rates before and after $t_0$ to calculate $\alpha$, and by measuring the total LEM positron rate along with a dedicated measurement for the uncorrelated positron rate, the tagging efficiency for individual muons can be extracted.

\subsection{\label{sec:LEMrate} LEM rate $r_\mathrm{LEM}$ }
Fig.~\ref{fig:decay_histo} shows a typical decay histogram from a weak transverse field \LEmuSR~measurement. The LEM rate $r_\mathrm{LEM}$, as indicated in Eq.~\ref{eq:lem}, is the total rate of tagged events for which a positron was detected by a positron counter. $r_\mathrm{LEM}$ can be extracted from a decay histogram by integrating the counts over the entire acquisition window of \SI{13}{\micro\second} across all positron counters, then dividing by the total acquisition time and normalizing to the proton beam current that generates the initial surface muon beam (up to \SI{2.4}{\milli\ampere} from the PSI High Intensity Proton Accelerator (HIPA)~\cite{grillenberger_high_2021}). 
A typical value for $r_\mathrm{LEM}$ is currently around \SI[per-mode=symbol]{1}{\kilo\hertz\per\milli\ampere}. With future improvements, such as an upgrade to the $\mu$E4 beamline, this rate is expected to be increased by a factor of 1.4 in 2026 \cite{2022_Zhou}.

\subsection{\label{sec:alpha} Background ratio $\alpha$}

\begin{figure}[t!]
	  \centering
	\includegraphics[width=0.99\columnwidth, trim={0 0 0 0},clip]{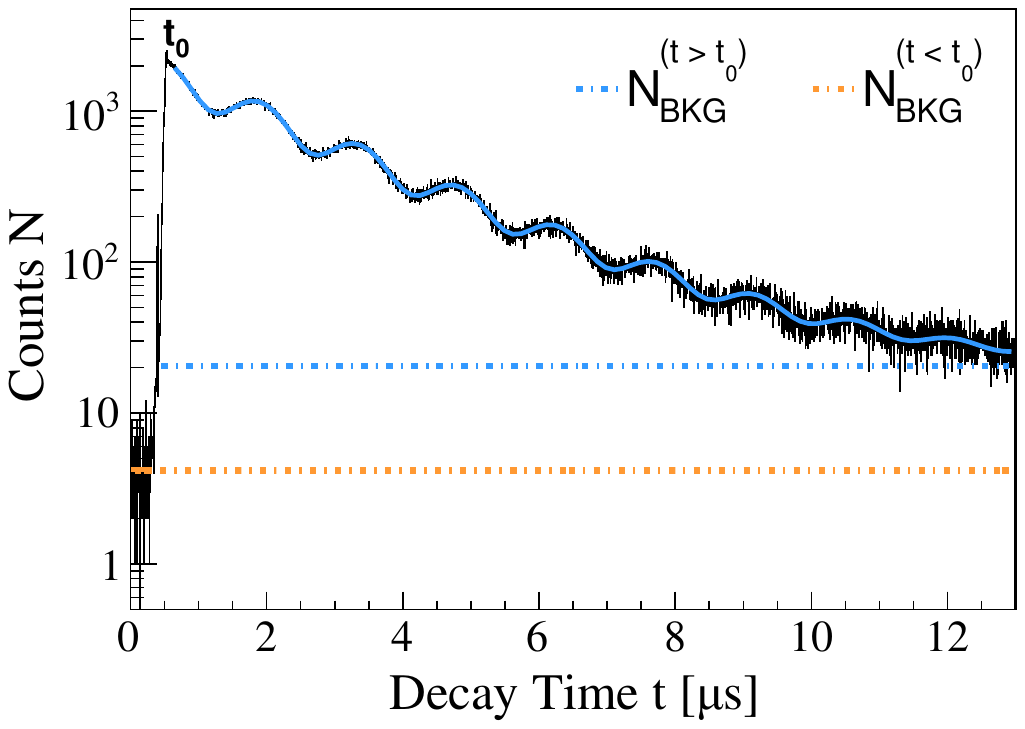} 
	  \caption[]{\label{fig:decay_histo}Typical decay histogram for a silver sample in a field of \SI{50}{\gauss}, measured at the LEM beamline with the positron detector PosL. The blue solid line represents the single-histogram fit described by Eq.~\ref{eqn:singlehist}, the blue dash-dotted line indicates the background level after t$_0$, while the orange dash-dotted line represents the background level before t$_0$.}
\end{figure}

\begin{figure*}[t!]
	  \centering
	\includegraphics[width=0.99\textwidth, trim={0 0 0 0},clip]{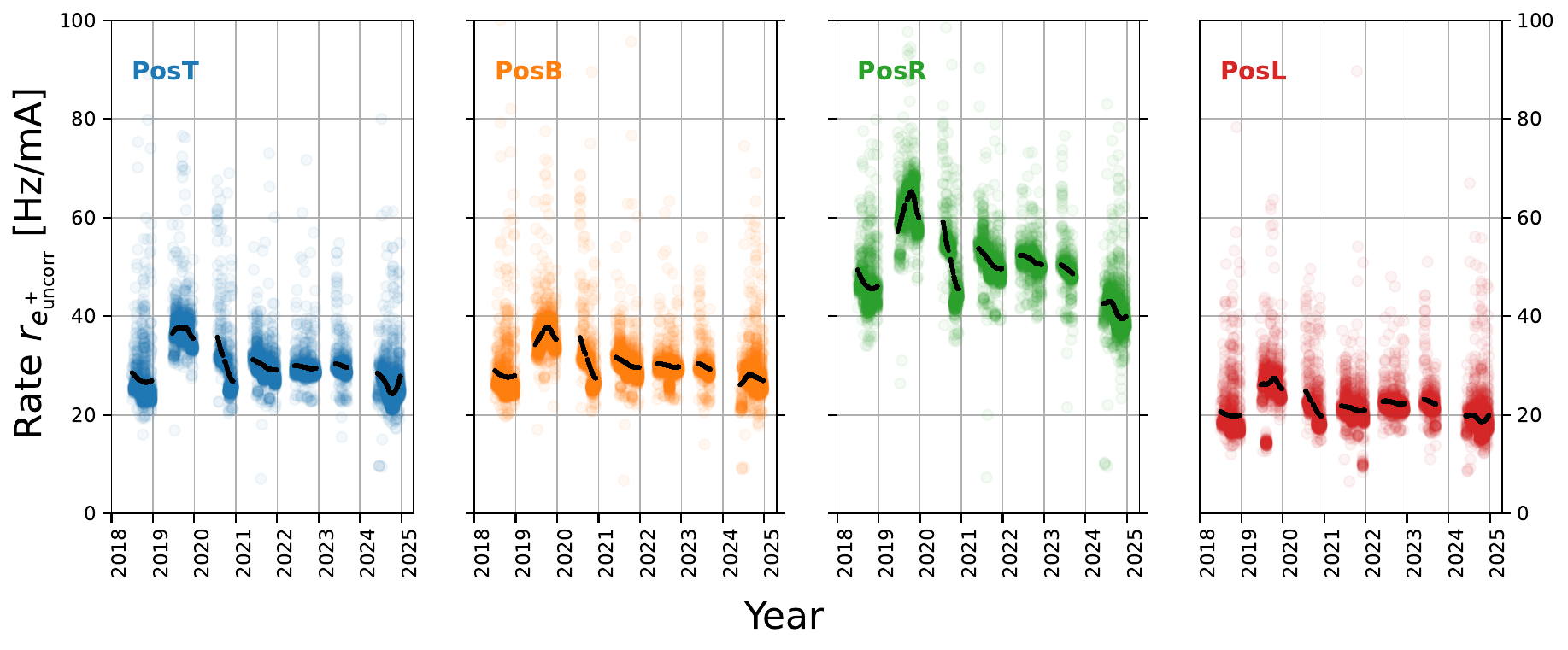} 
	  \caption[]{\label{fig:uncorr_BKG} Uncorrelated e$^+$ background rate $r_{e^+_\mathrm{uncorr}}$ at the LEM beamline for each detector segment (PosT, PosB, PosR, PosL), extracted for the years 2018 to 2024. The black curves, obtained via LOESS regression, are included as visual guides.}
\end{figure*}

$N_\mathrm{BKG}^{(t<t_0)}$ is extracted from the decay histogram by fitting a horizontal line. The fitting interval is limited to a maximum of \SI{0.3}{\micro\second}.
The fitting model used to extract $N_\mathrm{BKG}^{(t>t_0)}$ is as follows:
\begin{align}
    \label{eqn:singlehist}
    N(t>t_0) &= N_0 \overbrace{e^{-t/\tau_\mu}}^{\mu^+~decay}[1+\overbrace{A(t)}^\mathrm{decay~asymm.}]+N_\mathrm{BKG}^{(t>t_0)}\\
    \label{eqn:decay_asymm}
     A(t) &= A_0 \underbrace{e^{-\lambda t}}_\mathrm{relaxation} \underbrace{\cos(2\pi\nu t + \frac{\pi\phi}{180})}_\mathrm{oscillations}.
\end{align}
This fitting model represents a standard single-histogram fit (Eq.~\ref{eqn:singlehist}), which includes the total number of counts at $t_0$ ($N_0$), the muon decay with its natural lifetime ($\tau_\mu$ = \SI{2.1969811(22)}{\micro\second} \cite{Patrignani_2016}), the background level $N_\mathrm{BKG}^{(t>t_0)}$, and a term for the decay asymmetry $A(t)$ of the muon spin relaxation. 
The decay asymmetry $A(t)$ is modeled here as an exponentially damped oscillation (Eq.~\ref{eqn:decay_asymm}), including an initial asymmetry $A_0$, an exponential term describing the signal damping (characterized by $\lambda$), and an oscillatory cosine term defined by the frequency $\nu$ and phase $\phi$. Additional details on polarization functions of $\mu$SR data can be found in Refs.~\cite{1997_Dalmas, 2022_Blundell,2024_Amato}.
A fit was performed on an example dataset shown in Fig.~\ref{fig:decay_histo}, represented by the blue solid line. The fitting range is restricted to the interval between \SIrange{0.6}{13}{\micro\second}. In the fits, only the muon lifetime was treated as a fixed parameter.

This analysis enables the extraction of background levels before and after $t_0$ (see Fig.~\ref{fig:decay_histo}, dashed lines), which can then be used to calculate $\alpha$.

\subsection{\label{sec:UncorrBKG}Uncorrelated e$^+$ background $r_{e^+_\mathrm{uncorr}}$ }

During a \LEmuSR~measurement, the rate $r_{e^+_\mathrm{uncorr}}$ cannot be directly separated from the LEM rate, and therefore must be measured independently.
The simplest approach involves either turning off the high voltages of the electrostatic mirror or closing the BPVX valve (see Fig.~\ref{fig:lem}), which both prevents the slow $\mu^+$ beam from reaching the sample. 
During normal operation, the BPVX valve is closed automatically while warming up the sample to prevent sudden vacuum spikes in the moderator region. This provides ideal conditions to measure and record the uncorrelated e$^+$ background rate $r_{e^+_\mathrm{uncorr}}$. The $r_{e^+_\mathrm{uncorr}}$ values for the years 2018 to 2024 are shown in Fig.~\ref{fig:uncorr_BKG}, categorized by the positron counters (PosT, PosB, PosR and PosL). 
For visualization, the data were smoothed using LOESS regression \cite{1979_Cleveland,1988_Cleveland}, included in the scikit-misc Python package \cite{scikit-misc}.
As shown in Fig.~\ref{fig:uncorr_BKG}, PosT and PosB exhibit similar rates, while PosR shows nearly double the rate, and PosL is slightly lower. 
This discrepancy can be attributed to a gap in the lead shielding of the $\mu$SR spectrometer. This gap is caused by a vacuum connection going through the lead shield, which allows more scattered positrons to pass through towards the PosR detector. 

\section{\label{sec:results}Results and Discussion}

\begin{figure}[t!]
	  \centering
	\includegraphics[width=0.99\columnwidth, trim={0 0 0 0},clip]{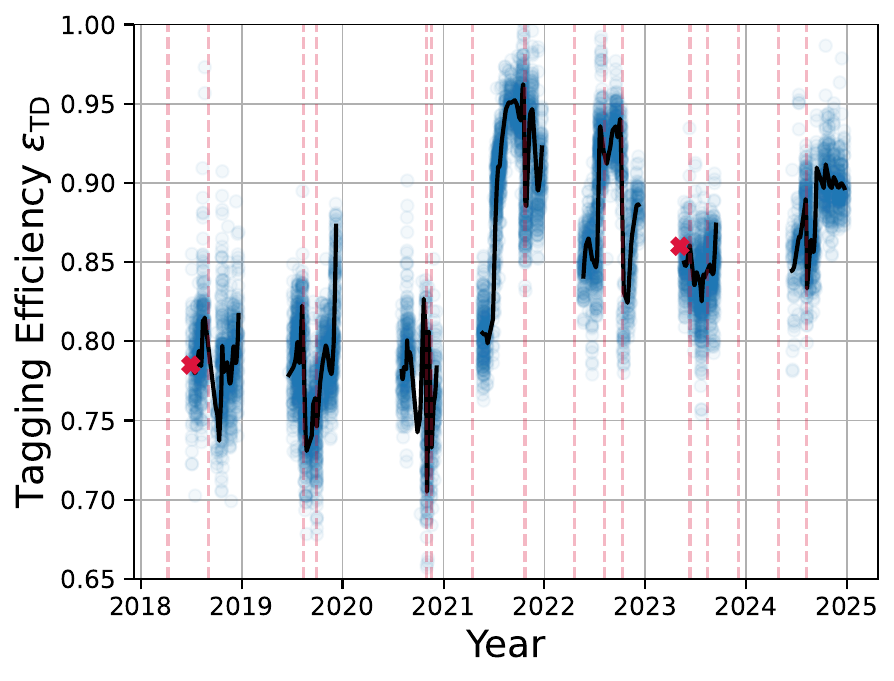} 
	  \caption[]{\label{fig:td_efficiency} Tagging efficiency, calculated using Eq.~\ref{eq:TDEff}, plotted for the years 2018 to 2024. The black curves, obtained via LOESS regression, are included as visual guides. The red dashed vertical lines mark the time of laser experiments. The red crosses represent reference measurements of the tagging efficiency performed using the method described in Ref.~\cite{2024_Janka}}.
\end{figure}

By extracting the total background levels across all four positron counters, as well as the LEM rate from all the standard weak transverse field (\SIrange{50}{100}{\gauss}) data collected at the LEM beamline between 2018 and 2024, and accounting for the varying uncorrelated background rate shown in Fig.~\ref{fig:uncorr_BKG}, the tagging efficiency $\epsilon_\mathrm{TD}$ was calculated and is presented in Fig.~\ref{fig:td_efficiency}.
As an initial validation, the presented method shows good agreement with tagging efficiencies determined using a dedicated second detector installed at the end of the beamline instead of a sample (red crosses in Fig.~\ref{fig:td_efficiency}, for method details see Ref.~\cite{2024_Janka}). 
The LOESS regression of the data (black curves in Fig.~\ref{fig:td_efficiency}) reveals that the tagging efficiency is rarely stable over time, exhibiting significant dips and fluctuations. 
The majority of the dips coincide with the red dashed lines in Fig.~\ref{fig:td_efficiency}, which mark the start time of a laser experiment conducted at the LEM beamline. In these experiments, the laser is directed through a view port at the mirror (Fig.~\ref{fig:lem}), passes through the carbon foil, and reaches the sample.
Previous work has demonstrated that laser irradiation removes contamination from the ultrathin carbon foil used in the tagging setup, thereby lowering the secondary electron yield and leading to a decrease in tagging efficiency \cite{2024_Janka}.
The observed increases in tagging efficiency are correlated with the type of samples being studied and the frequency of sample changes during experiments. 
Frequent (often daily) sample changes introduce significant contamination into the vacuum system, which gradually accumulates on the carbon foil. This buildup increases the effective foil thickness, thereby enhancing secondary electron emission and raising the tagging efficiency. However, this comes at the cost of degraded beam quality: an increased energy loss and straggling, a broader beam spot at the sample, and a reduced fraction of muons reaching the sample due to enhanced scattering or neutralization \cite{2024_Janka}. As a result, the quality of the low-energy $\mu$SR signal declines even as tagging efficiency rises. In particular, the observed asymmetry in $\mu$SR spectra may decrease, time and depth resolution worsen, and distortions from reflected low-energy muons (due to broader energy spread) may become more pronounced. 

In extreme cases, such as heavily outgassing samples or experimental incidents, the carbon foil's effective thickness can increase rapidly. 
An example of severe contamination occurred in July 2021, when a sample mounted on an oven setup, capable of reaching temperatures above \SI{500}{\degreeCelsius}, overheated due to a malfunctioning oven controller, ultimately burning the sample. 
This incident caused substantial deposition of contaminants on the carbon foil, as evidenced by the abrupt increase in tagging efficiency from 0.8 to 0.95. 
The original tagging efficiency of 0.8 was never fully restored, reaching only 0.85 following laser irradiation during the 2022 shutdown. 
In July 2022, a heavily outgassing sample (Methylammonium lead iodide, MAPI) was studied, which again increased the tagging efficiency to 0.93. Similar to the previous cases, only an extended laser experiment was able to restore the foil afterwards. 

Another valuable aspect that can be studied by monitoring the tagging efficiency is the effective coverage of the carbon foil. Examining the trend in tagging efficiency from 2018 to the end of 2020, it appears that the efficiency gradually decreases. A supporting indicator is the minimum tagging efficiency observed after laser irradiation; it decreased from approximately 0.75 in 2018 to 0.70 in 2020. This trend suggests that the aging of the carbon foil, possibly due to the formation of holes over time, may have contributed to a reduction in active area for secondary electron emission.

During the shutdown in 2023, the carbon foil was replaced with a thinner one, which underwent extensive laser irradiation to achieve optimal thinness, minimizing the energy loss and straggling, and thereby optimizing the beam spot size \cite{2024_Janka}. Comparing the minimum tagging efficiency of the old carbon foil (0.70 in 2020) with that of the new foil (0.83 in 2023), an increase is clearly visible. The improvement is likely due to better foil coverage and fewer defects or holes. 

In September 2024, another heavily outgassing sample, combined with being heated on the oven setup, again caused the carbon foil thickness to increase. This change went undetected for the remainder of the measurement campaign.  However, with the newly developed monitoring tool, such incidents can now be identified in real-time, allowing the foil parameters to be restored immediately via laser irradiation. This significantly improves the reproducibility of \LEmuSR~measurements and ensures more consistent experimental conditions.

\section{\label{sec:summary}Summary and Prospects}
We devise a new method to extract the tagging efficiency $\epsilon_\mathrm{TD}$ of the LEM beamline, a key parameter for achieving efficient and reproducible \LEmuSR~measurements, by comparing the background levels before and after the time $t_0$ in \LEmuSR~decay histograms.
Our results reveal significant variations in tagging efficiency over time, with increases attributed to sample-related outgassing and contamination, which lead to a greater effective thickness of the carbon foil. Similarly, decreases in efficiency are linked to laser experiments that remove contamination and reduce the foil's effective thickness.
This new monitoring method represents a valuable addition to the LEM beamline diagnostics, providing timely insights for determining when laser cleaning, or in extreme cases a replacement, of the carbon foil is required.
A major advantage of this method is that it requires no additional detectors or dedicated measurements, as the tagging efficiency can be extracted entirely as a byproduct of routine \LEmuSR~measurements. 
While developed for the LEM beamline, this method is broadly applicable to other beamlines that employ dual detection setups (e.g. in our case tagging detector and positron counters), provided that the background contributions differ at specific times, are well-characterized, and measurable. The most promising applications include other continuous muon beamlines capable of detecting individual muons, such as those at PSI \cite{himb}, TRIUMF \cite{triumf}, and the MuSIC beamline in Japan \cite{music}.


\begin{acknowledgments}
All the measurements have been performed at the Swiss Muon Source S$\mu$S, Paul Scherrer Institute, Villigen, Switzerland. This work is funded by the Swiss National Science Foundation under the grant number 220823 (GJ).
\end{acknowledgments}


\bibliography{apssamp}
\end{document}